\documentclass[11pt]{article}

\usepackage{hyperref}

\usepackage{amsmath}

\usepackage{graphicx}

\begin{document}

\title{Health risk modelling by transforming a multi-dimensional unknown distribution to a multi-dimensional Gaussian}

\author{Varun Kapoor\\
MPI-biophysical chemistry, G\"ottingen\\
vkapoor@gwdg.de, kapoorconsciousness@gmail.com
}
\maketitle

\begin{abstract}
The traditional approach of health risk modelling with multiple data sources proceeds via regression-based methods assuming a marginal distribution for the outcome variable. The data is collected for $N$ subjects over a $J$ time-period or from $J$ data sources. The response obtained from $i^{th}$ subject is $\vec{Y}_i=({Y}_{i1},\cdots, {Y}_{iJ})$. For $N$ subjects we obtain a $J$ dimensional joint distribution for the subjects. In this work we propose a novel approach of transforming any $J$ dimensional joint distribution to that of a $J$ dimensional Gaussian keeping the Shannon entropy constant. This is in stark contrast to the traditional approaches of assuming a marginal distribution for each $Y_{ij}$ by treating the $Y_{ij}'$s as independent observations. The said transformation is implemented in our computer package called ENTRA.          

\end{abstract}
\section{Introduction}

Information about the health outcomes in many epidemiological studies is obtained from multiple data sources or over a certain time-period with multiple observations. The multiple data sources provide multiple measures of the same underlying variable, measured on a similar scale. As an example $N$ adults are chosen for high-blood pressure study at the age of $18$ and are asked about their diet, smoking and drinking habits and their blood pressures are measured. The same subjects over the course of time are monitored again on the basis of the same variables choice as before. This illustrates the standard data collecting exercise to measure health risk. Once such a data is available we can begin the risk modelling to ascertain the factors which contribute towards high-blood pressure and those that contribute towards low-blood pressure. 

To put it mathematically, given $N$ subjects and $J$ data sources or time-points the data is collected as a $p$ dimensional vector of covariates, $\vec{X}_{ij}=({X}_{ij1},\cdots,{X}_{ijp})$ where $1<i<N$ and $1<j<J$. Given such a vector the outcome is reported as a variable $Y_{ij}$ for the $i^{th}$ subject and from the $j^{th}$ data source. Then we can construct a $J$ dimensional vector $\vec{Y}_i$ as $\vec{Y}_i=({Y}_{i1},\cdots, {Y}_{iJ})$. In a $J$ dimensional space the above vector for a given value of $i$ is a point. Since $1<i<N$ the total number of points in the $J$ dimensional space equal $N$. These $N$ points follow a certain distribution which is a-priori unknown. In the conventional analysis the outcome variables $Y_{ij}'$s are treated as independent variables and nothing is assumed about the joint distribution in the $J$ dimensional space. The assumption about the independence is not correct but as we will see in Section [\ref{tradition}] this does not affect the statistical analysis that we intend to carry out.\\ 

We propose a novel analysis tool to carry out health risk modelling by transforming the $J$ dimensional a-priori unknown density to that of a Gaussian density whilst keeping the Shannon entropy constant. To do so we transform the $N$ number of $J$ dimensional vectors in a basis set consisting of divergence-free vector fields. \cite{ward}. The condition that the basis set can only consist of divergence-free vector fields enforces the entropy conserving condition. Entropy conserving condition can be enforced by having volume preserving maps and our choice of the basis set represents a flow of incompressible fluid \cite{fluid,histentr} hence is a volume preserving map.

To determine the coefficients of these basis vectors such that the $J$ dimensional density is a Gaussian, Karplus theorem is used \cite{karplus,karptwo}. We will demonstrate in Section[\ref{vecs}] that how this theorem allows for determination of the basis coefficients in such a way that the $J$ dimensional density is transformed to a Gaussian.\\

The paper is divided as follows, in Section[\ref{tradition}] the traditional approach to model health risk is reviewed and a novel approach is proposed. In Section[\ref{Basis}] the construction of basis set consisting of high dimensional vector fields is shown. In Section[\ref{vecs}] the question of determining the coefficients of the basis set is settled. The algorithm and program structure is discussed in Section[\ref{algo}]. In Section[\ref{test}] a test case is computed to see how the transformation works in practice. Finally we conclude our work done so far and make proposals for the future work.

\section{Regression methods for multiple source outcomes}\label{tradition}
Consider a trial to model high-blood pressure with $N$ subjects monitored over $J$ time-period. The response obtained from the $i^{th}$ subject can be written as a $J$ dimensional vector as

\begin{equation}
\vec{Y}_i=({Y}_{i1},\cdots, {Y}_{iJ}).\label{master}
\end{equation}
Traditionally a joint distribution for $N$ subjects is not specified.  Instead a working generalized linear model (GLM) to describe the marginal distribution of $Y_{ij}$ as in Liang, Zeger (1986) \cite{zeger} is used,
\begin{equation}
f(Y_{ij}) = \exp\left[\frac{Y_{ij}\theta_{ij} - a(\theta_{ij})}{\phi} + b(Y_{ij},\theta_{ij})\right]\label{marginal}
\end{equation}

If the output $Y_{ij}$ is a binary random variable then the parameters for the above exponential family are 
\begin{equation}
\phi=1, a(\theta_{ij})=log[1+e^{\theta_{ij}}], \theta_{ij}=log\left[ \frac{\mu_{ij}}{1-\mu_{ij}}  \right], b(Y_{ij},\theta_{ij})=0
\end{equation}
The probability of a favorable outcome if $Y_{ij}$ is a binary random variable can be modelled via a Logit function as

\begin{equation}
\mathrm{Logit}(P[Y_{ij}=0 \arrowvert \vec{X}_{ij}]) = \vec{X}_{ij}\vec{\beta},\label{logit}
\end{equation}

where $Y_{ij}=0$ represents the favourable outcome \textit{i.e} low risk of high-blood pressure whereas $Y_{ij}=1$ corresponds to high-risk of high-blood pressure.\\
Given Eq[\ref{marginal}] the log-likelihood function can be written as
\begin{equation}
ln L = \sum_{i=1}^N\sum_{j=1}^J Y_{ij}\theta_{ij} - a(\theta_{ij}) 
\end{equation}
To determine the regression parameters $\vec{\beta}$ we differentiate the log-likelihood with respect to $\vec{\beta}$, this gives the following equation to estimate $\vec{\beta}'$s

\begin{equation}
\frac{\partial In L }{\partial \vec{\beta}}=\sum_{i=1}^N\sum_{j=1}^J \vec{X}_{ij}(Y_{ij}-\mu_{ij})=0.\label{assump}
\end{equation}

This is the traditional approach to determine the regression parameters that help us evaluate the factors which contribute towards high or low health risks given the data. Equation[\ref{assump}] was derived assuming that the outcome $Y_{ij}$ is a binary random variable, however similar equation can be derived when the outcome $Y_{ij}$ is of any other type.\\

In this approach it is assumed that all the $Y_{ij}'$s are independent observations. This assumption although not correct, yields the estimates for $\vec{\beta}$ which are valid but their variances are not. However using techniques such as empirical variance estimator valid standard errors can be obtained\cite{Huber}.
\subsection{A novel approach}
Having reviewed the traditional approach we now present our idea. As remarked earlier the joint distribution of the $N$ subjects is not specified in the $J$ dimensional space as this is an a-priori unknown. However if the unknown distribution is transformed to that of a $J$ dimensional Gaussian then a valid analysis tool can be developed. We have developed an algorithm and a program which does this transformation in an entropy conserving way \textit{i.e} the Shannon entropy is preserved during the transformation. \\

To preserve the Shannon entropy requires transformation of the high dimensional vectors in a basis set consisting of divergence-free vector fields as that represents a volume preserving map thereby preserving Shannon entropy. Such a construction of orthocomplete basis set is available in any dimensions \cite{ward}. We use this mathematical construct to transform then $N$ number of $J$ dimensional vectors. To determine the basis coefficients such that the $J$ dimensional density is transformed to that of a Gaussian Karplus theorem discussed in Section[\ref{vecs}] is invoked. Once the coefficients are determined the $J$ dimensional joint distribution of the subjects is a Gaussian having the same Shannon entropy as the starting distribution. This has been implemented in ENTRA.\\

Once the joint distribution for the $N$ subjects is known log-likelihood function can be written for such a distribution and the regression parameters determined thereby. In the next Section we show the construction of the divergence-free vector fields used in our program ENTRA. 

\section{$J$ dimensional divergence-free vector fields \label{Basis}}

A mathematically rigorous construction of divergence free smooth vector fields in any dimensions was provided in \cite{ward}. We use it to construct a basis set consisting of $J$ dimensional divergence free vector fields. To do so we define the following $J\times J$ matrix valued operator
\begin{equation}
\hat{\mathcal{O}}=-(I)_{J\times J}\nabla^2 +\nabla\nabla^T.\label{operator}
\end{equation}
Here the first term is a $J\times J$ dimensional Laplacian operator, $I$ being the $J\times J$ identity matrix and the second term consists of column and row vectors of the gradient operator in $J$ dimensions.

This operator acts on a smooth scalar function which we construct from a $J$ dimensional vector $\vec{x}$ as 

\begin{equation}
\phi_l\left(\arrowvert \arrowvert\vec{x}- \vec{x}_l     \arrowvert\arrowvert\right)= e^{- \arrowvert\arrowvert \vec{x}-\vec{x}_l\arrowvert\arrowvert^2/(2\sigma^2)  },\label{scalar}
\end{equation}

where the symbol $\arrowvert\arrowvert,\arrowvert\arrowvert$ is the Euclidean distance between two $J$ dimensional vectors. $\sigma$ is chosen to be $0.7\times \Delta x$ where $\Delta x$ is the spacing between the basis vectors. The vector $\vec{x}_l$ is chosen as a constant $\frac{l\Delta x}{2}$ where $l$ goes from $(-L,L)$  .

Now we define a matrix valued function by applying the operator in Eq[\ref{operator}] to the scalar field in Eq[\ref{scalar}] as

\begin{eqnarray}
&&\Phi_\sigma^l(\vec{x})_{J\times J}=\left\{-(I)_{J\times J}\nabla^2 +\nabla\nabla^T\right\}\phi_l\left(\arrowvert \arrowvert\vec{x}- \vec{x}_l    \arrowvert\arrowvert\right) \\ \nonumber &=&
 \left\{ \frac{J-1}{\sigma^2}-\frac{1}{\sigma^4}\arrowvert \arrowvert \vec{x}- \vec{x}_l\arrowvert\arrowvert^2 \right\}\left(I\right)_{J\times J} e^{- \arrowvert\arrowvert \vec{x}-\vec{x}_l\arrowvert\arrowvert^2/(2\sigma^2)  } \\ \nonumber &+&\left[\frac{1}{\sigma^4}(\vec{x}-\vec{x}_l)(\vec{x}-\vec{x}_l)^T     \right] e^{- \arrowvert\arrowvert \vec{x}-\vec{x}_l\arrowvert\arrowvert^2/(2\sigma^2)  }
\end{eqnarray}

It was proven rigorously that the columns of the above matrix consist of divergence free vector fields \cite{ward,thesis}. For a given choice of centre we therefore obtain $J\times J$ dimensional vector field. 
From the results in \cite{ward,thesis} $\nabla . \vec{v}_l^{J}=0$. \\

For a given centre $\vec{x}_l$ there are $J$ number of $J$ dimensional mutually orthogonal basis vectors. Hence for each centre we have a complete basis set. Due to such a construction the basis vectors enforce the divergence free condition strictly. Each vector has a unique coefficient $c_k$ attached to it,$1<k<J$.

In the next section we demonstrate this for a simple $2D$ case.

\subsection{2D case}

To demonstrate how the vector field looks like we take a simple 2D case and define the scalar operator in Eq[\ref{scalar}] as 

\begin{equation}
\phi_l\left(\arrowvert \arrowvert\vec{x}- \vec{0}     \arrowvert\arrowvert\right)=\frac{1}{\sqrt{2\pi\sigma^2}}e^{-(x^2+y^2)/{2\sigma^2}}
\end{equation}

The matrix valued function $\Phi_\sigma(\vec{x})_{3\times 3}$ becomes

\begin{equation}
\frac{1}{\sqrt{2\pi\sigma^2}}e^{-(x^2+y^2)/{2\sigma^2}}\left(
\begin{matrix}
-\frac{y^2}{\sigma^4}+\frac{1}{\sigma^2}&\frac{xy}{\sigma^4}  \\
\frac{xy}{\sigma^4}  & -\frac{x^2}{\sigma^4}+\frac{1}{\sigma^2} \end{matrix} \right)
\end{equation}

The vectors $\vec{v}_c^{1}$ and $\vec{v}_c^{2}$ defined from the columns of the above matrix as below are then divergence free,

\begin{equation}
\vec{v}_c^{1}=\frac{1}{\sqrt{2\pi\sigma^2}}c_0e^{-(x^2+y^2)/{2\sigma^2}}\left( \begin{matrix}
-\frac{y^2}{\sigma^4}+\frac{1}{\sigma^2}  \\ \frac{xy}{\sigma^4}  \end{matrix}\right)
\end{equation}

\begin{equation}
\vec{v}_c^{2}=\frac{1}{\sqrt{2\pi\sigma^2}}c_1e^{-(x^2+y^2)/{2\sigma^2}}\left( \begin{matrix}
\frac{xy}{\sigma^4}  \\ -\frac{x^2}{\sigma^4}+\frac{1}{\sigma^2} \end{matrix}\right)
\end{equation}

\begin{figure}
  \includegraphics[width=0.52\textwidth]{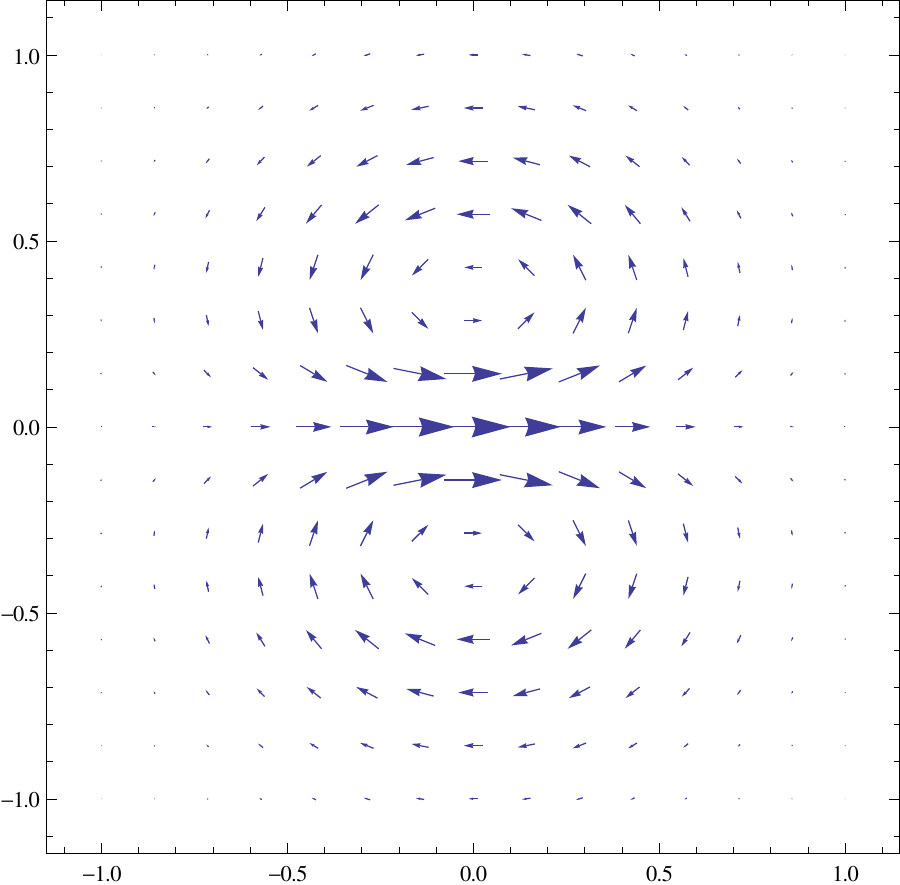} ;\
    \includegraphics[width=0.52\textwidth]{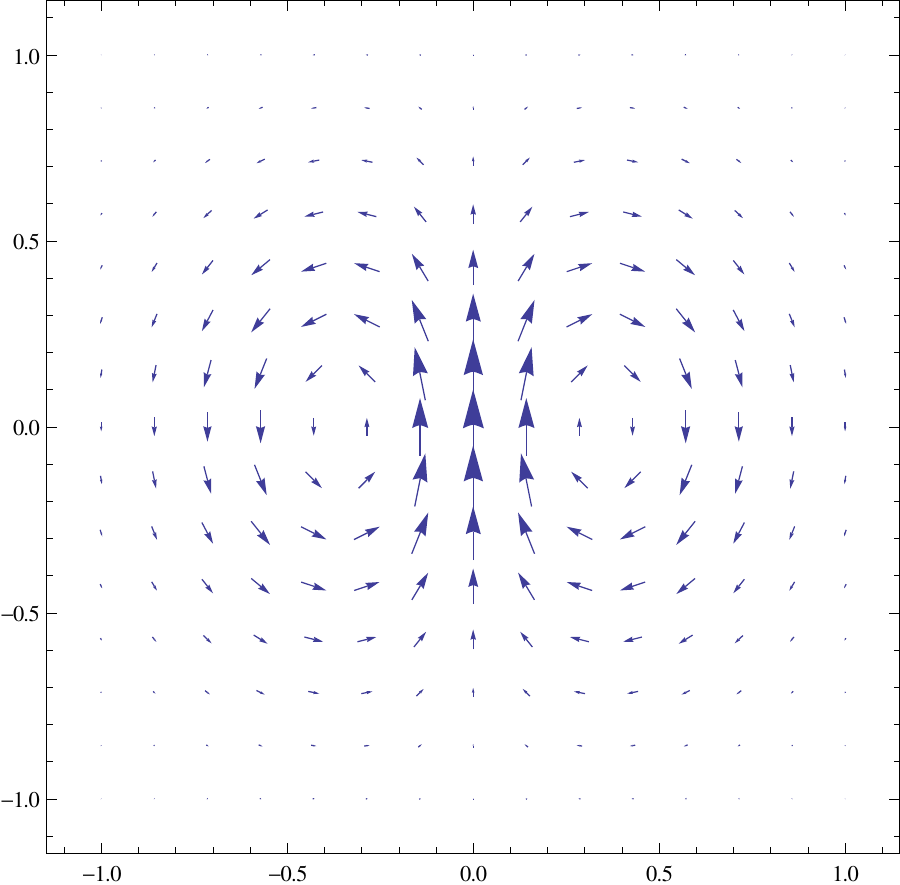}
	\caption{Plot of vector field  $\vec{v}_c^{1}$ and $\vec{v}_c^{2}$ respectively.	}\label{div}
\end{figure}

Any linear combination of these divergence free fields is also divergence free. We plot these fields in Fig.\ref{div} for $c_0=c_1=1$ and $\sigma=0.35$.

From the above plot we can see that for the $2D$ case we have two mutually orthogonal divergence free basis vectors which constitute the complete basis set in two dimensions. The result holds in general for any dimensions \cite{ward,thesis}.

\section{Vector Transformations}\label{vecs}
In the $J$ dimensional space of the outcome variable the $N$ subjects are represented by $N$ points. These $N$ points are arranged according to a certain density $\rho$. Karplus theorem states that for a given covariance Gaussian distribution maximizes entropy. Mathematically this can be written as

\begin{equation}
J[\rho]=S[\rho_G]-S[\rho]\geq 0.\label{basic}
\end{equation}
Here $\rho_G$ is the Gaussian density.

The covariance matrix is defined as
\begin{equation}
C=\langle (\vec{x}-\langle\vec{x}\rangle) (\vec{x}-\langle\vec{x}\rangle)^T \rangle.\label{covar}
\end{equation}
Here the symbol $\langle \rangle$ denotes the ensemble average over the $N$ subjects implying that $C$ is a 
$J\times J$ dimensional matrix. $\vec{x}$ denotes a $J$ dimensional vector or a point in the configuration space.\\
The equality sign in Eq[\ref{basic}] holds only if the underlying density distribution is a Gaussian.
Now we introduce the transformation $f$ that preserves the entropy, i.e, 

\begin{equation}
S[f(\rho)]=S[\rho].
\end{equation}

With this transformation we want to deform the density $\rho$ towards a Gaussian density, this then becomes the following minimization problem \cite{henn}

\begin{equation}
f_{\mathrm{min}}= \underset{f\in G}{\mathrm{min}}\ \left( J[f(\rho)]   \right).
\end{equation}

Here $G$ is the group of all the smooth entropy preserving transforms.
Since the transformation leaves the entropy unchanged and as the entropy of a Gaussian density is proportional to the determinant of the covariance matrix, to solve the above minimization problem we have to minimize the determinant of the covariance matrix of the Gaussian density. Since by Karplus theorem the covariance of $\rho_G$ is same as that of $\rho$, we can use the covariance in Eq[\ref{covar}] and write the above minimization problem as

\begin{equation}
f_{\mathrm{min}}=\underset{f\in G}{\mathrm{min}}\ \mathrm{det}C[f(\rho)].
\end{equation}

As a consequence of the above corollary if we have a basis set consisting of divergence free vector fields under which $J$ dimensional vectors for $N$ subjects are transformed, then the basis coefficients can be determined by minimizing the covariance matrix determinant of the transformed vectors with respect to the basis coefficients.


In the next section we describe the construction of the basis set consisting of divergence free smooth vector fields. This construction is implemented in the program package ENTRA, that I have developed.

\section{Algorithm and program structure}\label{algo}

Given $N$ subjects the outcome variable for each subject is a $J$ dimensional vector[\ref{master}]. We compute the covariance matrix Eq[\ref{covar}] which is a $J\times J$ dimensional matrix. The matrix is then diagonalized by an orthogonal transformation $T$. The covariance matrix $C$ can then be written as

\begin{equation}
C=T\Lambda T^T
\end{equation}
with $\Lambda$ being the eigenvalue matrix and columns of $T$ matrix being the eigenvectors of $C$.
We construct a $J \times N$ dimensional vector by appending all the $N$ number of $J$ dimensional vectors and label it as trajectory $\vec{Y}$. 
To transform the trajectory to principal coordinates where the mean of the trajectory is centered at $0$ we project the eigenvectors onto the trajectory to get mean-centered trajectory as

\begin{equation}
\vec{Y}^{'}=T(\vec{Y}-\langle\vec{Y}\rangle).
\end{equation}
For the trajectory $\vec{Y}^{'}$ we choose two $J$ dimensional vectors $\vec{Y}^{'}_ k$ and $\vec{Y}^{'}_ m$ and transform them in the vector field shown before as
\begin{equation}\vec{V}_ k=\vec{Y}^{'}_ k+\sum_ l^L\sum_i^J c_{i l} \Phi_{\sigma}^{il}(\vec{Y}^{'}_ k).\end{equation}
\begin{equation}\vec{V}_ m=\vec{Y}^{'}_ m+\sum_ l^L\sum_i^J c_{i l} \Phi_{\sigma}^{il}(\vec{Y}^{'}_ m).\end{equation}
The coefficients are chosen by minimizing the determinant of matrix $C$ with respect to the basis coefficients $c_ l$. The minimization is performed via conjugate gradients method \cite{cg}.
 
The process is repeated for all the $N$ number of $J$ dimensional vectors. This in the end yields a trajectory $\vec{V}$ which has the least determinant of the covariance matrix and the underlying density has the same Shannon entropy as that of our starting system.\\

\subsection{Program input}

The classes that build up the core of the program are ENTRA, trajectory and grid. Objects of class trajectory represent real-valued arrays. The arrays are stored as high dimensional vectors and the dimensionality of the vectors is defined by the grid class. The grid class also defines the number and spacing of the basis vectors. The methods provided by trajectory and grid classes allow to initialize the corresponding arrays, to manipulate them, and to store (load) them to (from) files. To start using the program few parameters have to be provided these are:\\
long nsources = J;\\
long nsubjects = N;\\
double deltx = spacing between basis functions;\\
long ngpsx = Number of basis functions=L;\\
long Max iter = Maximum iteration to find basis coefficients;\\

\section{Test Calculations}\label{test}

The goal of this example is to generate a trajectory having a random underlying density and then to transform it towards a trajectory having Gaussian underlying density. To generate a random trajectory we use the utility $Random$ of Eigen to generate random matrix of any dimension. The parameters chosen for this test are as follows:\\
long nsources=10 ;\\
long nsubjects=1000;\\
double deltx =0.05;\\
long ngpsx=80;\\
long Max\textunderscore iter=500;

\begin{figure}

\includegraphics[width=1.0\textwidth]{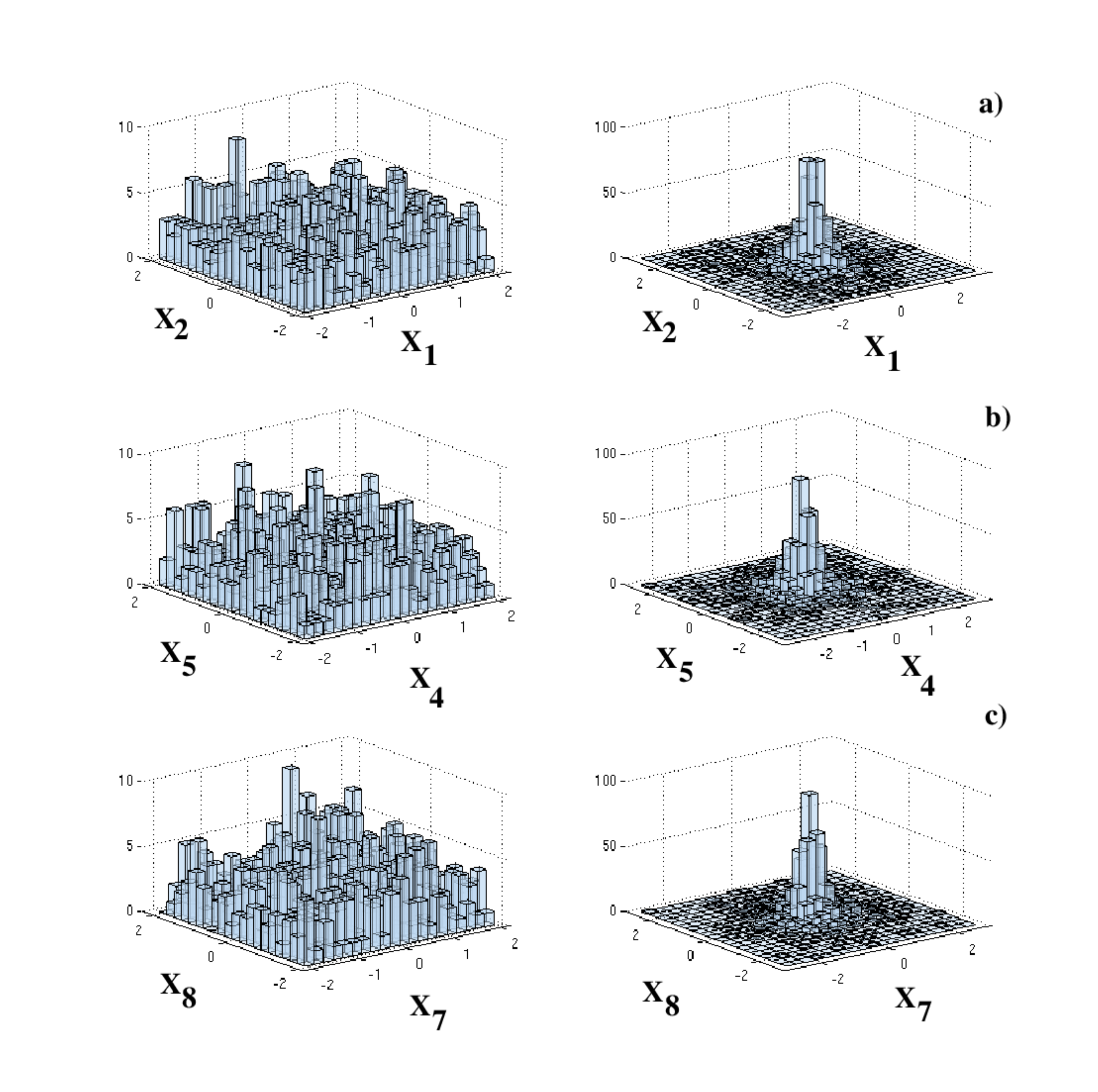}

\caption{2D Historgram plots for the original (left) and transformed (right) subspace. }\label{results}
 
\end{figure}

Results of the transformation (after a single iteration cycle) are shown in Fig.[\ref{results}]. In this example we transform a $30$ dimensional vector having $1000$ configurations. The data is along these $30$ axises which are labelled as $(X_1,X_2\cdots X_{30})$. Here $X_p=Y_{ip}$, \textit{i.e} $p^{th}$ component of the $J$ dimensional vector. We plot two dimensional subspace of original $30$ dimensional space along different axises as labelled in Fig.[\ref{results}].

To prove that the transformation is indeed entropy conserving, we computed subspace entropy via histogramming for the 2D subspaces shown in Fig.[\ref{results}] for the original and the transformed subspaces.\\

The Matlab code file along with the output to compute entropy of subspaces is provided here:\\
*******************************************************************************************************\\ 
Matlab File to estimate Entropy via Histogramming for the plot in Fig.[\ref{results}]a\\
*******************************************************************************************************\\
load simulation\_examplehist.dat

X1=simulation\_examplehist(:,8);

X2=simulation\_examplehist(:,9);

X3=simulation\_examplehist(:,11);

X4=simulation\_examplehist(:,12);

X11=[X1;X3];
X12=[X2;X4];

X = [X11,X12];

plotmatrix(X);

defaultn=500;  

error(nargchk(1, 2, nargin));         
if nargin $< 2 $  
   n =defaultn ;                           
end 

X = double(X);                        
Xh = hist(X(:), n);                   
Xh = Xh / sum(Xh(:));                 

i = find(Xh);           

h = -sum(Xh(i) .* log2(Xh(i)));       

InitialEntropy =h;

display(InitialEntropy);

Y1=simulation\_examplehist(:,2);

Y2=simulation\_examplehist(:,3);

Y3=simulation\_examplehist(:,5);

Y4=simulation\_examplehist(:,6);

Y11=[Y1;Y3];
Y12=[Y2;Y4];

Y=[Y11,Y12];

plotmatrix(Y);

error(nargchk(1, 2, nargin));         
if nargin $< 2 $  
   n = defaultn;                           
end 

Y = double(Y);                        
Yh = hist(Y(:), n);                   
Yh = Yh / sum(Yh(:));                 

i = find(Yh);           

htwo = -sum(Yh(i) .* log2(Yh(i)));       

TransformedEntropy=htwo;
display(TransformedEntropy);

EntropyDifference=InitialEntropy-TransformedEntropy;

display(EntropyDifference);

***************************************************************************************** \\
Result of the above file\\
*****************************************************************************************\\
$>>$ run simpleentropy
InitialEntropy =

    8.7636

TransformedEntropy =

    7.5625

EntropyDifference =

    $ 1.2011$\\
*****************************************************************************************************

Result for the plot Fig.[\ref{results}]b\\

$>>$ run simpleentropy

InitialEntropy =

   8.7770

TransformedEntropy =

   7.6632

EntropyDifference =

  $ 1.1138$\\
*****************************************************************************************************

Result for the plot Fig.[\ref{results}]c\\

$>>$ run simpleentropy

InitialEntropy =

   8.7754

TransformedEntropy =

   7.6483

EntropyDifference =

     $1.1271$

Two dimensional histogram plots for initial and transformed configurations are shown in Fig.[\ref{results}].

\begin{figure}
\includegraphics[width=1.0\textwidth]{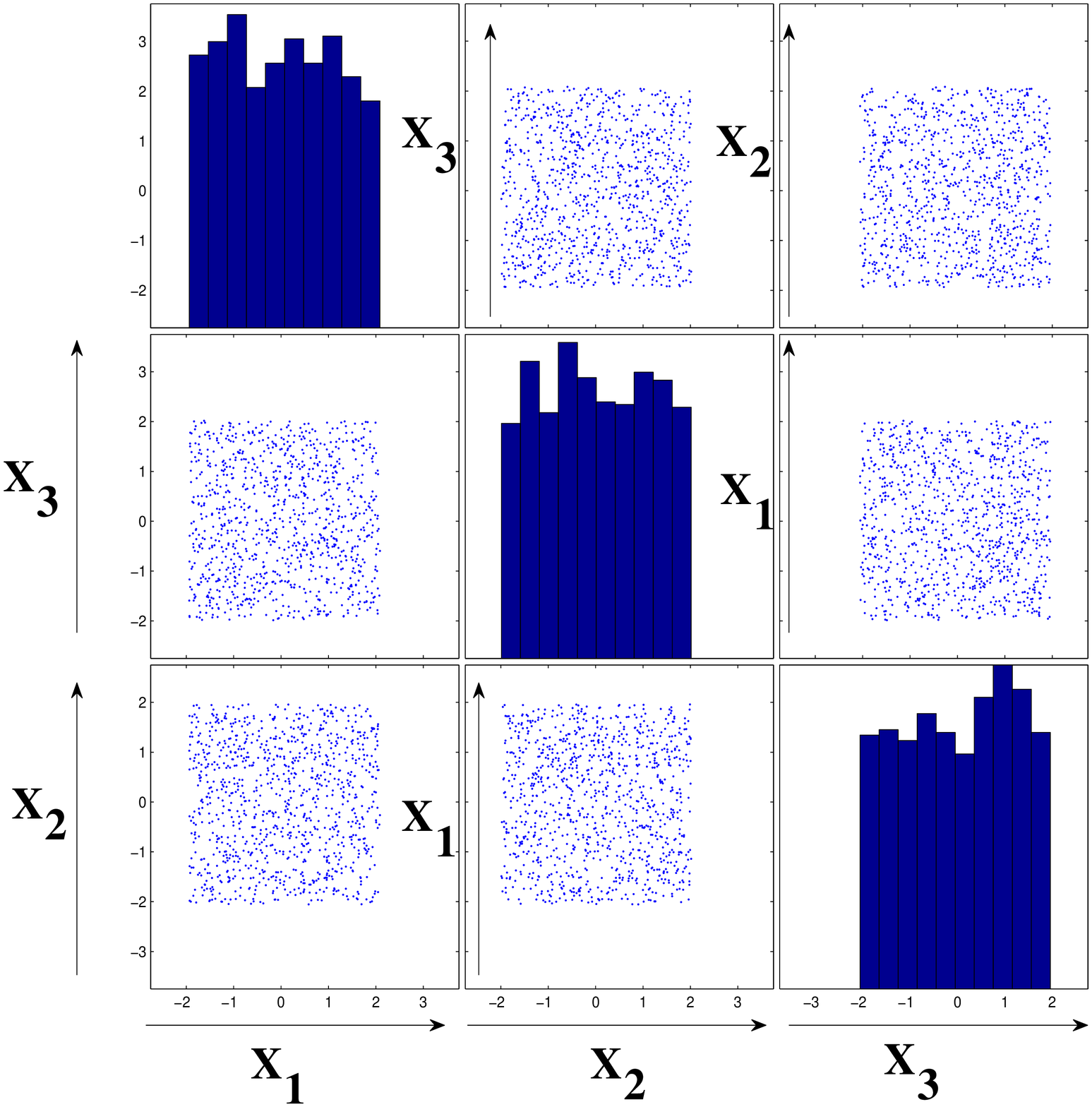}
\caption{3D Historgram plots for the original subspace along three axises $(X_1,X_2,X_3)$. More details in the text }\label{threedorg}
 \end{figure}

\begin{figure}
\includegraphics[width=1.0\textwidth]{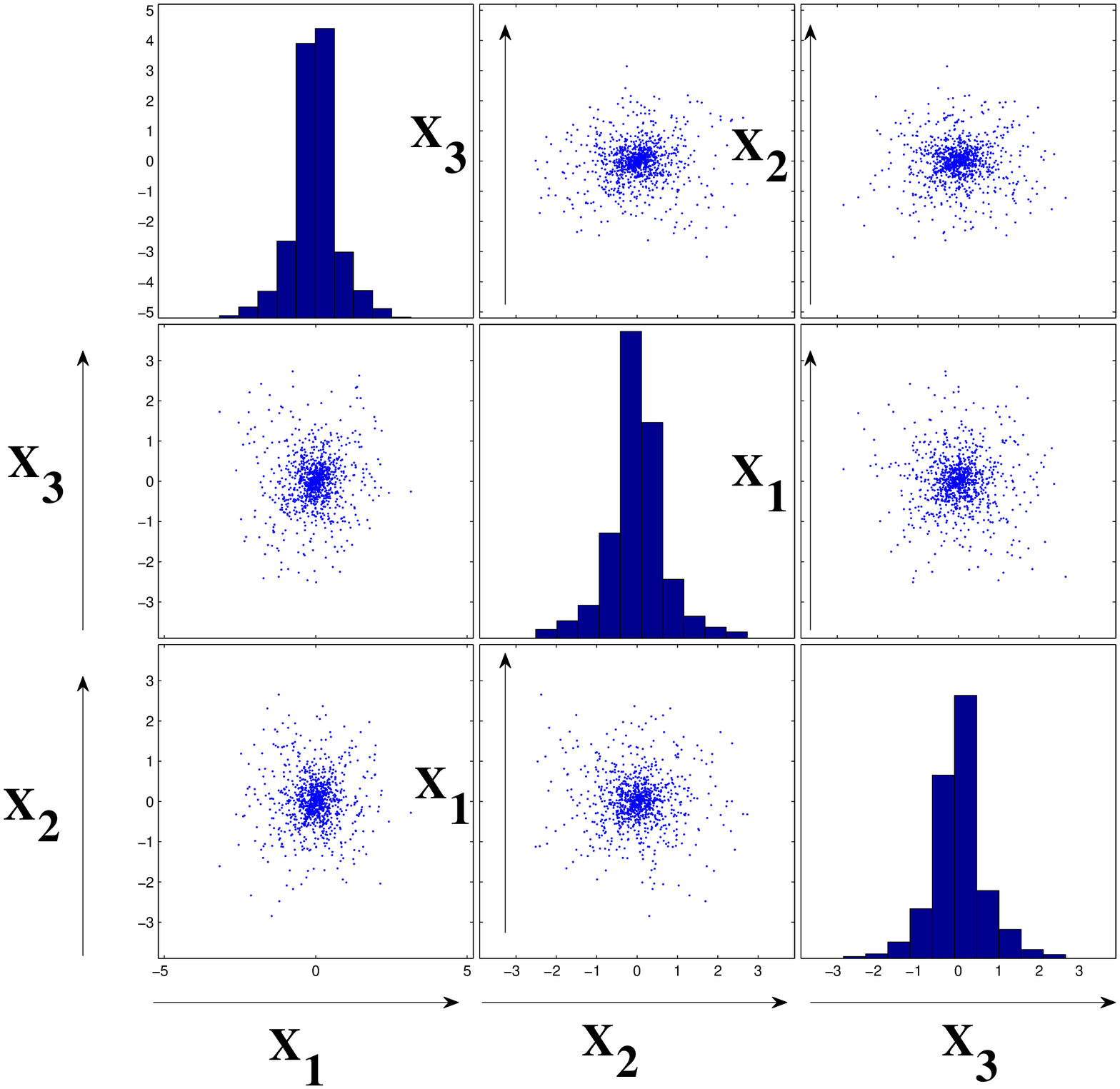}
\caption{3D Scatter and Historgram plots for the original subspace along three axises $(X_1,X_2,X_3)$. More details in the text }\label{threedtrans}
 \end{figure}

Now we can also compute entropy in higher dimension via histogramming. As the data is 30 dimensional we now look at the data along first three axises $(X_1,X_2,X_3)$ as seen in Fig.[\ref{threedorg}], in this plot we plot the data along one axis versus another as labelled along with histogram along each axis. We also plot the same plot for the transformed data as seen in Fig.[\ref{threedtrans}]. We compute the entropy of the three dimensional data using a Matlab code similar to above and its results are shown here:

*****************************************************************************************************

Result for computation of entropy for 3D data

$>>$ run highdimentropy

InitialEntropy =

   8.8220

TransformedEntropy =

   7.5756

EntropyDifference =

     $1.2464$


These results imply that in just a single iteration cycle the unknown configuration space density is transformed to a Gaussian density with entropy being conserved approximately. Clearly $1000$ points are not sufficient to get an accurate entropy estimate and more statistics is required. This will form part of the work to be done.

\newpage
\section{Summary and Outlook}
ENTRA package has been presented. The aim of this package is to do data transformation on high dimensional data sets as found in epidemiology. With this transformation the underlying high dimensional density function is transformed that to a high dimensional Gaussian and due to the nice properties associated with a Gaussian distribution the further data analysis can be accomplished easier than before.

The following major points need to be addressed which will form the main body of the work to be done at the institute, they are: 

1) The appropriate choice of the basis set vectors. The number of basis vectors depend on the data and need to be appropriately estimated beforehand. How exactly that can be achieved needs to be determined. \\

2) Building an example with enough statistics to be able to prove that the entropy conservation is maintained to a high degree of accuracy.\\

3) Furthermore developing full file support for epidemiologists to enable them to load their data and get the transformed data.\\ 

4) Also, developing complete regression analysis to estimate conditional probabilities of the likes in Equation[\ref{logit}] in the ecosystem of ENTRA.












\end{document}